\documentclass{aastex}
\usepackage{emulateapj5}

\def\ltsima{$\; \buildrel < \over \sim \;$}
\def\simlt{\lower.5ex\hbox{\ltsima}}
\def\gtsima{$\; \buildrel > \over \sim \;$}
\def\simgt{\lower.5ex\hbox{\gtsima}}
\received{}
\accepted{}
\journalid{}{}
\articleid{}{}
\slugcomment{To appear in ApJ}
\lefthead{}
\righthead{}

\shorttitle{M87 Temperature Structure}
\shortauthors{S.Molendi}

\begin{document}
\title{On the Temperature Structure of M87}
\author{Silvano Molendi}
\affil{Istituto di Astrofisica Spaziale e Fisica Cosmica - IASF, CNR, 
Sezione di Milano, \\
via Bassini 15, I-20133 Milano, Italy}
\email{silvano@mi.iasf.cnr.it}
\begin{abstract}
We revisit the XMM-Newton observation of M87 focusing our attention on 
the temperature structure. We find that spectra for most regions of M87 can 
be adequately fit by single temperature models. Only in a few regions, 
which are cospatial 
with the E and SW  radio arms,  we find evidence of a second temperature. 
The cooler component (kT $\sim$ 0.8-1 keV) fills a small volume compared to 
the hotter component (kT $\sim$ 1.6-2.5 keV), it is confined to the 
radio arms rather than being associated with the potential
well of the central cD and is probably structured in blobs with typical 
sizes smaller than a few 100 pc. Thermal conduction must be suppressed 
for the cool blobs to survive 
in the hotter ambient gas. Since the cool gas is observed only in those regions 
of M87 where we have  evidence of radio halos our results favor models in which
magnetic fields play a role in suppressing  heat conduction.   
The entropy of the cool blobs is in general smaller than that of the hot 
phase gas thus cool blobs cannot originate from adiabatic evolution of 
hot phase gas entrained by buoyant radio bubbles, as suggested by Churazov et al. 
(2001). An exploration of alternative origins for the cool gas leads to 
unsatisfactory results.

\end{abstract}


\section {Introduction}
 
M87 is the giant elliptical galaxy at the center of 
the Virgo cluster. It is amongst  the most studied objects
in the sky. Observations at radio and X-ray wavelengths, over the past 
few years, have allowed us to improve substantially our understanding of
the physical processes in this source.
The ROSAT analysis of M87 (B\"ohringer et al. 1995) has shown  
excess emission correlated with the radio structure and that this emission 
originates from gas cooler than the surrounding intracluster medium.
A detailed radio map of M87 obtained with the VLA at 90 cm (Owen et al. 2000)
provided evidence for a very complex radio structure. The authors identified 
two ``bubbles'' of synchrotron emission located respectively E and SW some 
20-30 kpc from the nucleus. 
The bubbles are most likely supplied with energy and relativistic plasma 
which come ultimately from the active nucleus and the inner jet of M87.
Largely on the basis of the above X-ray and radio observations Churazov 
et al. (2001) have developed a model for M87. These authors propose that
the bubbles discovered by Owen et al. (2000) buoyantly rise through the cluster 
atmosphere. During their rise the bubbles uplift relatively cool X-ray 
emitting gas from the central regions to larger distances. Since the 
entrained gas is cooler than the ambient gas, the radio bubbles are 
trailed by cool elongated X-ray features.
From the analysis of the XMM-Newton observation of M87, Belsole et al. (2001) 
have shown that cospatially with the radio bubbles  the X-ray emission can 
be modeled as the sum of two single temperature components, with 
temperatures of about 2 and 1 keV respectively. 
In a recent paper based on a Chandra observation of M87, Young et al. (2002) 
have confirmed the results reported by Belsole et al. (2001).  

In this Letter we revisit the XMM-Newton observation, and also make some 
use of a Chandra observation of M87. We perform a detailed spectroscopic 
analysis with the aim of better constraining the temperature structure of 
M87 and consequently gain a deeper understanding of the physical processes 
at work in this source.       
The remainder of the paper is organized as follows. In Section 2 we briefly 
review the XMM-Newton observation and data preparation. In Section 3 we describe 
the spectral modeling of the M87 data. In Section 4 we focus on the cool 
component emission. Finally in Section 5 we summarize our main results.

%
%
%
%

\section {Observations and Data Preparation}

We use XMM-Newton EPIC data from the PV observation 
of M87/Virgo.
Details on the observation, as well as results from the 
analysis of this object, have already been  published in various papers 
including B\"ohringer et al. (2001), Belsole et al. (2001), 
Molendi \& Pizzolato (2001),  Molendi \& Gastaldello (2001),
Gastaldello \& Molendi (2002) and Matsushita et al (2002).  

We have obtained calibrated event files for the MOS1, MOS2 and pn
cameras with SASv5.2. 
In this Letter we use pn data only for the imaging analysis described in 
Section 4, spectral analysis has been performed on MOS1 and MOS2 data. 
At the time of writing, the calibration of the MOS detector in the 
relevant energy range (0.4-4.0 keV) is better than that of the pn detector, 
moreover the superior spectral resolution of the MOS with respect to pn allows 
a better modeling of the line rich spectra observed in M87.
 
Data were manually screened to remove any remaining bright 
pixels or hot column. 
There have been no soft-proton flares during the 
M87 observation and consequently no rejection for 
time periods with high background has been applied.
We have divided the M87 observation in 8 concentric annuli
centered on the emission peak. Each ring has been further divided
into sectors for a total of 107 regions. 

Spectra have been accumulated for MOS1 and MOS2 independently.
The Lockman hole observations have been used for the background. 
Periods in which the background is 
increased by soft-proton flares have been excluded using an 
intensity filter; we rejected all events accumulated  when 
the count rate exceeds 15 cts/100s in the $[10-12]$ 
keV band.
Background spectra have been accumulated 
from the same detector regions as the source spectra.
The vignetting correction has been applied to the spectra 
rather than to the effective areas, similarly
to what has been done by other authors who 
have analyzed EPIC data (e.g. Arnaud et al. 2001).
Spectral fits were performed in the 0.4-4.0 keV band.
Data below 0.4 keV were excluded to avoid residual calibration 
problems in the MOS response matrices at soft energies.
Data above 4 keV were excluded because above this energy the 
spectra show a substantial contamination from hotter gas emitted 
farther out in the cluster, on the same line of sight. 
      
\section {Spectral Modeling}

All spectral fitting has been performed using version 
11.0.1 of the XSPEC package. All models discussed below
include a multiplicative component to account for the
galactic absorption on the line of sight of M87.
The column density is always fixed at a value of
1.8$\times10^{20}$ cm$^{-2}$, which has been determined  
from a detailed 21 cm  measurement in the direction of M87 
(Lieu et al. 1996). The above value is somewhat smaller than
the one derived by Stark et al. (1992), 2.5$\times10^{20}$ cm$^{-2}$,
obtained at a lower spatial resolution. We note that such small
variations in column density do not affect in any substantial
way our results.

We have first compared our data with a single temperature model
(vmekal model in XSPEC), 
which allows to fit separately individual metal abundances. 
This model has 14 free parameters: the temperature $T$, the
normalization and the abundance of  $N$, $O$, $Ne$, $Na$, $Mg$,
$Al$, $Si$, $S$, $Ar$, $Ca$, $Fe$ and $Ni$, which are all expressed 
in solar units. 

As shown in Gastaldello \& Molendi (2002) using the apec rather
than the mekal model yields very similar results as far as the temperature
and the normalization are concerned. Since temperatures
and  normalizations are the quantities we 
are interested in we have not run the apec model on our data.

In Figure 1 we show the temperature map obtained from the 
single temperature model run on the data. 
The map shows that in general the temperature is smaller at smaller radii.
However the most interesting result is the presence of cooler regions 
which extend from the core in the E and SW directions. We will refer to 
these regions from now on as the E and SW ``arms''. 
Our findings are in agreement with previous analysis of the XMM M87 observation 
by Belsole et al. (2001) and Matsushita et al. (2002).

We have then compared our data with a two temperature model
(vmekal + vmekal model in XSPEC). This model has two 
free parameters more than the single temperature model: 
the temperature and the normalization of the second component.
Since we are interested in the temperature structure we have chosen to
link  the metal abundance of  the second component 
to that of the first component. From tests on sample spectra we have found that,
with the exception of Fe,
allowing the metal abundances of the two components to vary independently 
results in little difference in the normalizations  and temperatures 
of the best fitting model.
In the case of  Fe there is a correlation between the normalization and the Fe abundance 
for  the cooler component, because the spectral range where this component contributes 
most significantly to the observed spectra is around 0.8-1.0 keV where the cooler 
component emission is dominated by the Fe-L shell line blend. 
Tests on sample spectra show that
the correlation between  normalization and Fe abundance of the cooler component 
is such that, within the 90\% confidence contours, 
the former can be reduced up to 30-50\% if the latter 
is allowed to increase by a similar amount. 

To establish if spectra are better represented by a 2 temperature
model than by a 1 temperature model we have made use of the F-test. In 
Figure 2 we show the value of the probability, associated to the F-test, 
for the improvement to be statistically significant. Quite interestingly 
it is only for the two arms extending from the core in the  E and SW 
directions that we have evidence of more than 1 temperature. For 
all other regions  spectra are consistent with being emitted by a
single-phase gas.

One possible way to test the robustness of our characterization of the cooler component 
is to take the spectrum from a region where we have evidence of two temperatures
and subtract from this the spectrum of a nearby region at the same angular distance 
from the center of M87 showing no evidence of the cool component. Since, 
as detailed later on in this section, the hotter
component does not feature strong azimuthal variations, we expect 
that the spectrum resulting from the above subtraction should be well  
represented  by a single temperature model
with temperature and normalization similar to those found for the cool component.
We have conducted the above test on 4 sectors, some in the SW and others in the E 
arm, in all cases we find that the resulting 
spectra can be well fitted by a single temperature model with temperature and 
normalization  differing no more than 10\% from those previously derived from the 
two temperature fits of the same regions.   

We note that the presence of a second temperature component in the E 
and SW arms was already discussed in the Matsushita et al. (2002)
and Belsole et al. (2002) paper; the statistical test we conduct here
firmly establishes that this second thermal component is found 
in the SW and E arms and nowhere else at a comparable statistical 
significance level.

In general the reduced $\chi^2$ we obtain from our 2 temperature fits 
indicate that this model provides a fair description of the data.
There are three regions for which the probability associated to the 
$\chi^2$ is smaller than 0.01, visual inspection of the residuals
shows that in all three cases an important contribution to the  total $\chi^2$
comes from two spectral ranges: the first around 1.1-1.3 keV and the 
second around 1.7-1.9 keV range. Inspection of other spectra,
where the fits are formally acceptable, indicates that residuals 
in the above energy ranges are rather common. For the  1.1-1.3 keV
range the residuals may reflect an incomplete modeling of the Fe-L 
lines in the mekal code while for the 1.7-1.9 keV they are probably 
due to an insufficient modeling of the Si edge fine structure 
in the detector quantum efficiency. 

We have fitted spectra showing evidence of a second thermal
component with a 3 temperature model (vmekal + vmekal + vmekal model in 
XSPEC) and with a 2 temperature plus cooling-flow model (vmekal + vmekal + 
vmcflow in XSPEC). By making use of the F-test  we find that the 
introduction of a third temperature or of a cooling-flow component 
does not provide a statistically significant improvement to the fit.
Thus, from an observational point of view, our results on the temperature 
structure may be summarized as follows.
For most regions the gas can be well described by
a single temperature model. For the two arms extending respectively 
E and SW the spectra can be adequately described by a two temperature 
model. 
The lack of improvement in the fit when going to models more 
complicated than 2 temperatures does not mean that the gas is strictly 
composed of  two phases and there is no emission at all at any other 
temperature. The correct way of interpreting our result is that the bulk of the 
emission is coming from gas with a temperature close to either one of the 
two temperatures found in our best fits. Emission from any other component is
relatively small. To quantify the latter statement we have performed the following 
exercise on the sector with bounding radii 2$^\prime$ and 3$^\prime$ and bounding angles 
150$^o$ and 180$^o$ (angles are counted starting from West and going counter clockwise), 
which is characterized by a particularly intense cool 
component. We have added a third temperature component to our 
two temperature model and  frozen the temperatures of the first 2 components
as well as the normalization of the hotter component to the best fitting values found 
from the 2T fit. 
We have then run fits stepping  the temperature of the third component
from 0.5 to 5 keV. For each fit we have derived 
the 90\% confidence upper limit for the normalization of the third component. 
In Figure 3 we show the ratio of the 90\% upper limit  for the normalization 
of the third component over the cool component normalization as a function of 
the temperature of the third component. 
The plot nicely illustrates
that a third component can contribute substantially with respect to the cool component 
measured in the two temperature fit, only if its temperature 
is larger than $\sim 0.8$ keV, emission from 
gas at temperatures smaller than 0.6 keV has an associated normalization that is 
no more than 2\% of that of the cool component. 
Some emission from gas with temperatures
in the range between 1.7 and 5 keV cannot be ruled out, the intensity of this 
emission can be relatively large if compared to that of the cool component, however
it is much smaller (a few percent) when compared to that of the hot component.   


In Figure 4 we report the ``normalized'' emission integrals (EI),
i.e. the emission integrals, as derived by spectral fitting, 
divided by the area of the region over which they have been measured, 
as a function of radius (for more details see the Figure caption). 
For those sectors where we have evidence of 
two temperatures, we report the normalized EI of the hot and cool components
while for all other sectors we plot the normalized EI obtained from the 
one temperature fit. Similarly in Figure 5 we report the temperatures
as  a function of radius, showing both hot and cool component temperatures for 
two temperature regions and the temperature obtained from the 1T fits for
all other regions. We note that neither the normalized EIs nor the temperatures
for the hot or single phase component show evidence of a large scatter
at any given radius, indicating that no strong azimuthal gradients are present.
Moreover for those annuli where we have 1 and 2 temperature 
 measurements we find that the temperature and normalized EI of the hot 
component are not clearly separated from the temperature and normalized EI 
of the single temperature fits, as is to be expected if the hot component
observed in the 2 temperature regions is essentially the same we measure in 
the 1 temperature regions.  Thus the picture emerging from our analysis 
is that of a 2 temperature structure, with the hot component that is 
distributed in a regular and symmetric fashion and a cool component that 
is observed only in the SW and E arms. It is to this latter component 
that we now turn our attention.

\section{The cool component}

In a number of papers based on ASCA observations various authors (e.g. 
Ikebe et al. 1996, Makishima et al. 2001) have discussed 2 temperature 
models as an alternative to the multi-phase cooling-flow model preferred by 
other authors. 
In the two temperature model presented by the above authors the higher 
and lower temperature components are  associated
respectively to gas in the potential well of the cluster and of the cD
galaxy. The potential would thus be characterized by a sort of
hierarchical structure.
From Figure 2 it is evident that the lower temperature component 
has a highly asymmetric distribution which is reminiscent of 
the radio emission (cfr. Owen et al. 2000) and is not associated
to the cD which has a much more regular structure. Thus the EPIC
observation provides evidence against a model in which 
the lower temperature component is associated
to gas in the potential well of the cD galaxy.

We now wish to further investigate the nature of the cooler component, 
and more specifically to determine  the volume occupied by this component 
with respect to that occupied by the hot component.
For a given region this can be expressed as  $V_c \sim EI_c / n_c^2$, 
where $V_c$ is the volume occupied by the cool component and $EI_c$ and 
$n_c$ are respectively the emission integral and gas density of the 
cool component. Note that the above estimate is valid under the assumption that 
the cool component is confined to a scale in projection comparable to the 
observed extent of the arms within the plane of the sky. 
To compute $V_c$ we need an estimate of $n_c$, which can be obtained
by assuming that the hot and cool components are in pressure equilibrium with 
one another.  

There are a number of reasons that favor pressure equilibrium between the 
various  components:  
1) the X-ray data does not show evidence of shocks anywhere 
in M87, which would be expected in the presence of substantial pressure jumps 
and/or supersonic motions;  
2) radio observations (Owen et al. 2000) indicate that  
the radio bubbles are most likely not overpressurized with respect to the  
X-ray gas;
3) if strong pressure gradients were present between the hotter and cooler  
X-ray emitting gas and or the radio plasma they would  be reduced on a few  
sound crossing timescales which, assuming a typical size of a few kpc would 
be smaller than $10^7$ yr. 
Finally we note that, as far as our calculations are concerned, we do not 
require exact pressure equilibrium. Indeed as we shall see below our conclusions 
would not change if the pressure of the hot and cool components were within a factor 
of 2 of each other. 

Under the assumption of pressure equilibrium the volume occupied by the 
cool component can be expressed as $V_c \sim EI_c / n_h^2 \cdot kT_c^2/kT_h^2$, 
where  $kT_c$ and $kT_h$ are respectively the temperature of the cool
and hot components and $n_h$ is the gas density for the hot component.
The ratio between the volumes occupied by the cool and hot 
components may then be estimated in two different ways. In the first case 
we assume that, as for the cool component, 
$V_h \sim EI_h / n_h^2$, where $V_h$ and $EI_h $ are respectively the 
volume and the emission integral of the hot component.
In this case the ratio between the volumes occupied by the cool and hot 
components is simply 
$$V_c / V_h = EI_c / EI_h   \Big ( {kT_c \over kT_h }\Big )^2. \eqno(1)$$
Alternatively the hot component density can be recovered by deprojecting 
the M87 data and the volume occupied by the hot component can be 
estimated by assuming that the linear extent of a given region in projection 
is similar to that in the plane of the sky, i.e. $V_h = A^{3/2}$, where $A$ is
the area from which the spectrum has been extracted.
We have estimated the $V_c / V_h$ ratio for all regions where we have evidence of
a statistically significant cooler component applying both methods.
From direct application of eq. (1) we find that $V_c / V_h $ varies between a few 
times $10^{-3}$ and a few times  $10^{-2}$. 
For the second method the adopted deprojected $n_h$ profile has been
derived from Matsushita et al. (2002) and from our own deprojection
analysis (Pizzolato et al. in prep.), the two profiles are in 
good agreement with one another and so are the derived $V_c / V_h$
ratios. Estimates of $V_c / V_h $ based on the deprojected 
$n_h$ method are about a  factor of two larger than those obtained from 
direct application of eq. (1). 
The difference between the two estimates is due to the fact that
in the first case the emission integral for the hot component, $EI_h$,
overestimates $V_h \cdot n_h^2$ because it contains contributions 
from gas farther out in the cluster observed on the same line of sight.
This is borne out by estimates of the relative contribution 
to $EI_h$ from gas in a given spherical shell and from the overlaying gas 
observed on the same line of sight. 
Given the approximations we have made to compute  $V_c/V_h$, and the
difference in values returned by our two methods,  it is 
fair to say that our measurements are good within a factor of $\sim$ 2.

As shown in Figure 4 and  already discussed in Section 3, for those annuli 
where we have 1 and 2 temperature sectors, the normalized EIs for the hot component 
in 2 temperature sectors are similar to the normalized EIs for the single phase 
gas in one temperature sectors.
This implies that the SW and E arms do not contain substantially less hot 
phase gas than regions which do not feature strong radio emission and are located
at similar radial distances from the cluster core. 
If, as we have argued above, the radio emitting plasma is in pressure equilibrium
with the thermal gas, the above result implies then the filling factor of the radio 
plasma must be small. 
Owen et al. (2000), from the analysis of an image  of M87 at 90 cm, 
find that the radio halo is highly filamentary and speculate that 
the interfilament region could be dominated by the thermal plasma, 
our findings indicate that this may well be the case.

Our analysis shows that the volume occupied by the cooler
phase is, for all regions, much smaller than the volume occupied by the
hot phase.
Moreover the size of the individual structures, possibly blobs,
producing the cool emission, must be considerably smaller than the
size of the regions for which we have accumulated spectra.
To place tighter constraints on the size of these structures we have
produced an emission integral ratio map according to the following procedure.
We have produced two images: a ``soft'' image, S, extracting events from the 
0.8-1.0 keV band, where the ratio of the cool component to the hot component 
is largest and a ``hard'' image, H, extracting events from the 0.5-0.7 keV 
and 1.1-4.0 keV bands where, on the contrary, the hot  dominates over the 
cool emission. 
Assuming a cool component temperature of  0.85 keV and a hot component 
temperature of 1.7 keV, we have expressed  the ratio of the soft component 
to the hot component in the soft band as a function of the 
counts in the soft and hard bands.
We have then converted this in a cool over hot component emission
integral ratio and reported the resulting map in Figure 6. As can be seen,
the cool component emission is concentrated in the SW and E arms. 
Moreover the emission integral of the cool component is always between 0.01 and
0.2 of that of the hot component, roughly corresponding to a $V_c / V_h $ 
ratio in the range 0.0025 to 0.07 if we use eq. (1) and 0.005 to 0.14 if 
we correct for the factor of 2 found when comparing results obtained from 
eq. (1) to those obtained using deprojected values for $n_h$.
 
The obvious implication is that, even when pushing 
the resolution of the EPIC image to its limits  and perhaps a little beyond 
(the pixels in our images are $10\arcmin\arcmin\times 10\arcmin\arcmin $ ) 
we still do not resolve the cool component blobs.
To further constrain
the size of the cool component structures we have produced a new 
emission integral ratio map using a Chandra
observation of M87. The observation was carried out with the S3 chip 
in the aim point between the 29th and 30th of July 2000, the total effective 
exposure time, after minor cleaning for flares, is $\sim$35.3 ks, data 
was reprocessed using version 2.2.1  of CIAO and 2.10 of the calibration
database.
Thanks to the better spatial resolution of Chandra and high surface brightness 
of M87 we can employ $4\arcmin\arcmin\times 4\arcmin\arcmin$ pixels 
corresponding to 300 pc $\times$ 300 pc. The map has been 
produced following the same procedure adopted for the EPIC data. 
The Chandra map, see Figure 7, yields emission integral ratios 
similar to those found from the EPIC map, indicating that in most 
regions the size of the cool blobs is smaller than the resolution of the 
Chandra image. There is however one region  5$\times$2 pixels in size, 
localized $\sim 10 \arcmin\arcmin$ N of the nucleus, where the ratio has an 
unusually high value of $\sim 0.7$. 
In this region we may be observing a cool component blob with a linear 
size of $\sim $ 1 kpc. 
Our assessments of the typical sizes of blobs are ofcourse rather crude,
and are only meant as order of magnitude estimates. More detailed estimates
would probably require comparisons with simulations in three dimensions.

The survival of small cool blobs embedded in a hotter medium 
brings us to the role of thermal conduction. In the absence of any
inhibition, the timescale over which the blobs will evaporate from
contact with the hotter gas is very short, assuming a typical size 
of 300 pc and  density of $3\times 10^{-2}$ cm$^{-3}$, the conduction 
timescale is a few times $10^5$ yr (note that, even though we consider
relatively small sizes, we do not expect conduction to be saturated 
as the mean free path for electrons is roughly ten times smaller).  
Thus, either blobs are continuously generated by an unknown
mechanism or they survive thanks to  substantial suppression of thermal
conduction.
It is hard to envisage a mechanism capable of continuously generating 
cool blobs for at least two reasons, firstly
because the thermal conduction timescale is shorter than other timescales,
such as the sonic or the cooling timescales on which such a mechanism
may operate; secondly because, as we shall discuss in more detail later, the 
entropy of the cool blobs is smaller than that of the hotter gas from 
which they would presumably be formed.
Thus the most likely scenario is one in which thermal conduction is
suppressed. Assuming that the age of the blobs is equal or larger than 
the timescale over which the radio bubbles rise through the cluster atmosphere
(a few times 10$^{7}$ yr) we estimate that conduction must be suppressed by a 
factor of 100 or more. 
For many years it has been recognized that 
heat conduction might be suppressed in the ICM (e.g. Binney and Cowie 1981), 
and many authors (e.g. Chandran et al. 1999) have invoked magnetic fields as a possible 
means to achieve suppression. 
It is therefore quite remarkable that we find evidence 
of suppression only in those regions of M87 where we know magnetic field are 
present because we see radio halos. 

In Figure 8 we show the entropy $S$, defined as $S\equiv T/n^{2/3}$, 
for the cooler and hotter components as a function of the radius. 
The density is estimated from a spectral deprojection of the data (Pizzolato 
et al. in prep.), while the temperature is the emission weighted temperature
measured directly from spectral fits, using the density profile presented
in Matsushita et al. (2002) produces very similar results.
As can be seen, the entropy of the cool phase is virtually everywhere 
smaller than  the entropy of the hot phase. This brings us to the rather 
important question of how the cool blobs where formed.

In a recent paper Churazov et al. (2001) have developed a detailed model 
to explain the radio and X-ray structure observed in M87. 
The authors envisage a scenario in which radio bubbles rise  subsonically 
through the cluster atmosphere. As the bubbles rise,
they capture ambient gas, dragging it upwards,  during its upward motion 
the entrained gas is expected to evolve adiabatically. 
The above model was constructed to satisfy, amongst others, constraints coming 
from X-ray observations carried out with the ROSAT satellite.
It is therefore important to test whether the model can live up to the better
quality XMM-Newton EPIC and Chandra ACIS data. 
The model fails to explain two fundamental observational facts.
Firstly, the model predicts that as the blobs  rise their temperature 
drops adiabatically, while from Figure 5 we observe that the blob temperatures
are remarkably contained everywhere between 0.8 and 1.0 keV, with 
no indication of a temperature drop with increasing radius.   
Secondly,  since the cooler gas entropy is virtually everywhere smaller than the 
hotter gas entropy (see Fig. 8), the cooler gas cannot result from  adiabatic 
expansion of the hotter gas entrained by the radio bubbles.
Even assuming a conservative indetermination of a factor 2 on our relative
entropy estimates we can allow for an adiabatic evolution only of those 
blobs currently located at radii larger than $\simeq 2^{\prime}$, blobs at 
smaller distances from the core would remain unexplained.
  Note also that, since the timescale on which the 
bubbles rise, a few times $10^{7}$ years, is smaller than the cooling 
timescale of the gas, a few times $10^{8}$ years, the entropy 
of the entrained gas cannot have been reduced through cooling. 

The lack of a  temperature gradient for the cool gas argues against  
a long adiabatic rise of the blobs through the cluster atmosphere.
As suggested by Nulsen et al. (2001), the entrainment of the cool blobs
from the radio bubbles may be unstable, thus the blobs could be captured and then
released after having risen only a short way. If conduction is suppressed by magnetic
fields associated to the bubble than the release of the blob from the bubble could
be rapidly followed by heating of the blob to the ambient gas temperature, this
could explain the clear separation between the cool and the hot gas 
temperatures.
   
Perhaps the most puzzling issue is the low entropy of the cool blobs.
As already pointed out, this argues against the cool blobs originating 
from the ambient gas. 
One possible solution is to assume that, as envisaged by Fabian et al. (2001),
metals are distributed in a highly inhomogeneous way in the ICM. 
If, for example, about 1\% percent of the gas had a metallicity of 50, in solar units,
 and the rest of the gas had a metallicity of 1/2, again in solar units, than 
the metal rich gas would cool at a rate more than 20 times faster than the metal poor
gas. Outside the region dominated by radio bubbles thermal conduction would 
maintain the metal rich blobs at the same temperature of the metal poor gas.
In the radio bubbles, where magnetic fields suppress conduction, the metal rich
gas would cool on timescales comparable or even shorter  than those on which the radio 
bubbles rise through the cluster atmosphere. In this scenario the cool blobs 
are metal rich lumps which are kept at a roughly constant temperature due to a 
balance between radiative cooling and suppressed thermal conduction. Indeed, 
even if conduction
is suppressed,  cooling of the blobs increases 
the gradient between the cool and the hot phase making conduction more efficient.
The entropy problem would also be solved because, as the metal rich lumps 
cool, they  radiate away entropy.  Unfortunately this rather attractive scenario
does not do well when tested against the EPIC data. When fitting some of the 
spectra of the 2 temperature regions with a 2 temperature model allowing the
Fe abundance of the two components to vary independently,  we find that
the Fe abundance of the cool component cannot be larger than 3-4 in solar units,
even when using  99\% confidence intervals. 

Another alternative to explain the low entropy of the cool blobs is to  assume 
that they have been generated by heating of cooler gas, perhaps through 
interaction with the radio lobes. 
However detailed studies at longer wavelengths 
have shown that the mass of cold gas present in M87 is orders of magnitudes
smaller than the mass of the cool phase gas which we observe in X-rays.
For instance Sparks et al. (1993) estimate a total mass of the order of 
$10^{5}-10^{7} M_\odot$ in the form of ionized gas in dusty filaments,
while Edge (2001) derives an upper limit of $\sim 1.5\times 10^{8}  M_\odot$
for any molecular gas, this is to be compared with a total mass of about
$4\times 10^{9} M_\odot$ for the cool component, which we estimate from our 
measurements.
A possible way out would be to assume that the gas has all been heated
up, as might be expected if the core of M87 had recently undergone a 
starburst phase.
If this were to be the case, we would expect to observe an enhancement
in the abundance of heavy element. As discussed in Gastaldello and Molendi
(2002) and  in Canizares et al. (1982) we do indeed have evidence 
of an O overabundance in the core of M87,
however a simple estimate of the mass of the starburst required to explain 
the O excess yields  $4-5 \times 10^{8}$ M$_\odot$, which is an order of 
magnitude smaller than the mass of the cool gas we observe at X-ray 
wavelengths. 

\section {Summary}

We have performed an analysis of M87 using  XMM-Newton EPIC  and Chandra 
ACIS data. Our main findings may be summarized as follows.
\begin{itemize}

\item
  Spectra for most regions of M87 can be adequately fit by single temperature
   models. It is only for a few regions, which are cospatial with the E and SW 
  radio arms, that we find evidence of a second temperature. Fitting more 
  complicated
  spectral models to these regions (i.e. including a third temperature or 
  a cooling-flow  component) does not improve the quality of the fits.  

\item
  The lack of X-ray ``holes'' at the location of the SW and E radio arms 
  indicates that the radio plasma has a small filling factor. As already 
  suggested by Owen et al. (2000) the radio structures are most likely highly 
  filamented with the interfilamentary regions dominated by the hot phase 
  thermal plasma.   
  
\item
  The cool component fills a small volume compared to the hot component, it is 
   probably structured in blobs with typical linear sizes smaller than a 
  few 100 pc. 

\item
  Thermal conduction must be suppressed for these blobs to survive in the 
  hotter ambient gas. 
   Since the cool gas is observed only in those regions of M87 where we have 
   evidence of radio halos our results provide important evidence that magnetic 
   fields could be responsible for the suppression of heat conduction.

\item

   The model proposed by Churazov et al. (2001) is found to be in contradiction 
   with our observational results in two major aspects. 
   Firstly, it predicts that  as the blobs  rise their temperature 
    drops adiabatically, while we find 
   no indication of a temperature drop with increasing radius.   
   Secondly, the entropy of the cool blobs is in general smaller than that 
   of the hot phase gas, thus cool blobs cannot originate from adiabatic 
   evolution of  hot phase gas entrained by buoyant radio bubbles, as suggested 
   by Churazov et al. (2001).

\item
  We have investigated two alternative solutions to solve the puzzle of 
  the low entropy of the cool blobs. The first solution requires that the blobs,
  which originate from higher entropy ambient gas, rapidly radiate away their 
  excess entropy thanks to their high metal content. However detailed 
  spectral modeling of the EPIC data does not allow for a very high metal 
  content of
  the cool blobs. The second assumes the cool gas  originates from heating 
  of cold gas. However, current estimates of the mass of cold gas are 
  orders of magnitude smaller than the mass of cool gas we observe 
  $\sim 4\times 10^{9} M_\odot$.
  Thus,  at the present time,  we are not capable of providing a convincing 
  origin for the cool gas we observe in M87.

\end{itemize}

\acknowledgements
It is a pleasure to acknowledge fruitful discussions with Fabio Pizzolato, Fabio
Gastaldello, Simona Ghizzardi, Stefano Ettori, Gianfranco Brunetti and Sabrina
De Grandi. S. De Grandi is also thanked for reduction of the Chandra data.
An anonymous Referee is thanked for providing helpful comments.

\clearpage


\clearpage

\begin{figure}
\epsscale{1.0}
\plotone{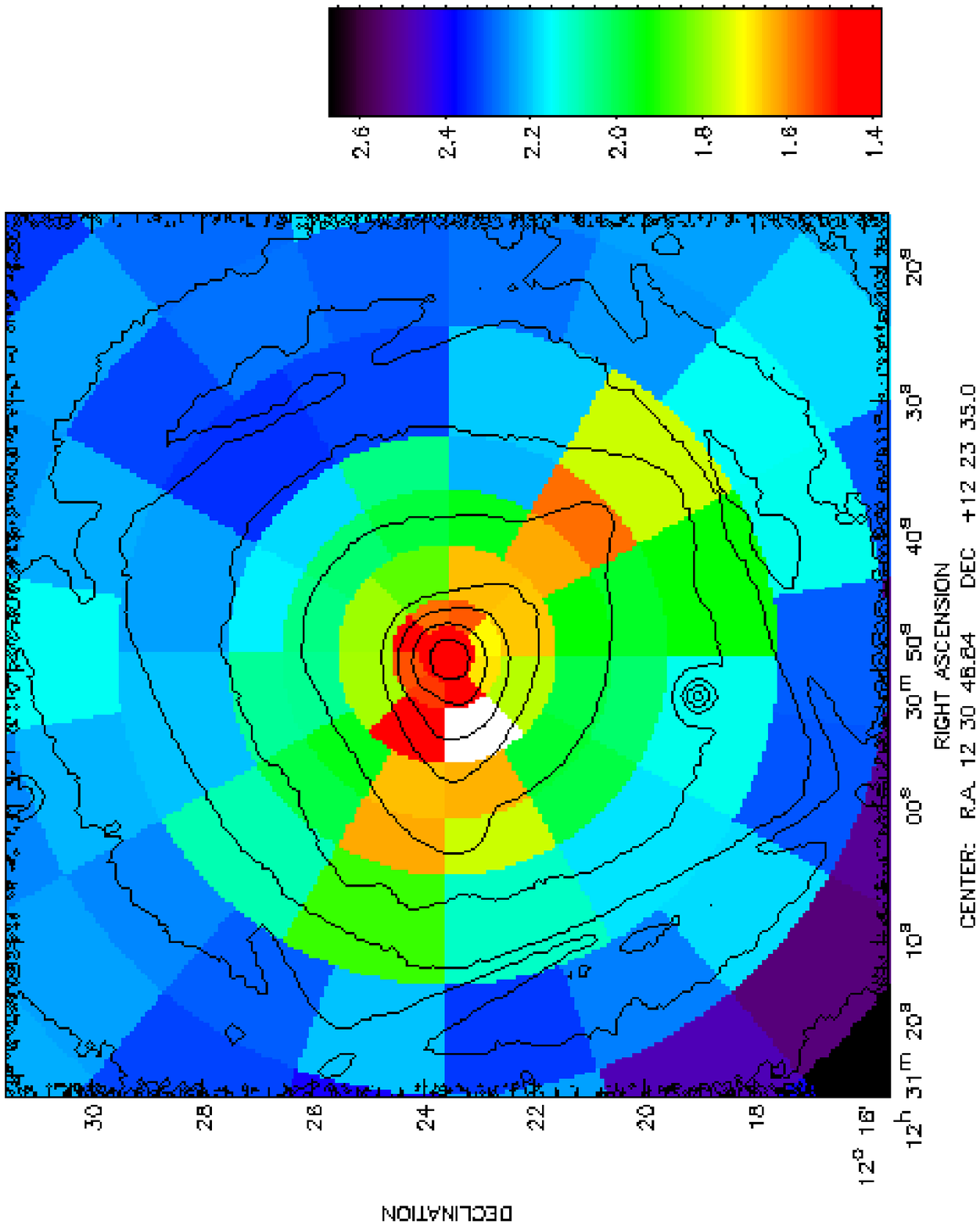}
\caption
{Temperature map, obtained from the 1 temperature run, overlaid on intensity
contours. The color palette on the right shows the correspondence
between colors and temperature expressed in keV. The white patch 
corresponds to a value slightly below 1.4 keV}
\end{figure}
\clearpage

\begin{figure}
\epsscale{1.0}
\plotone{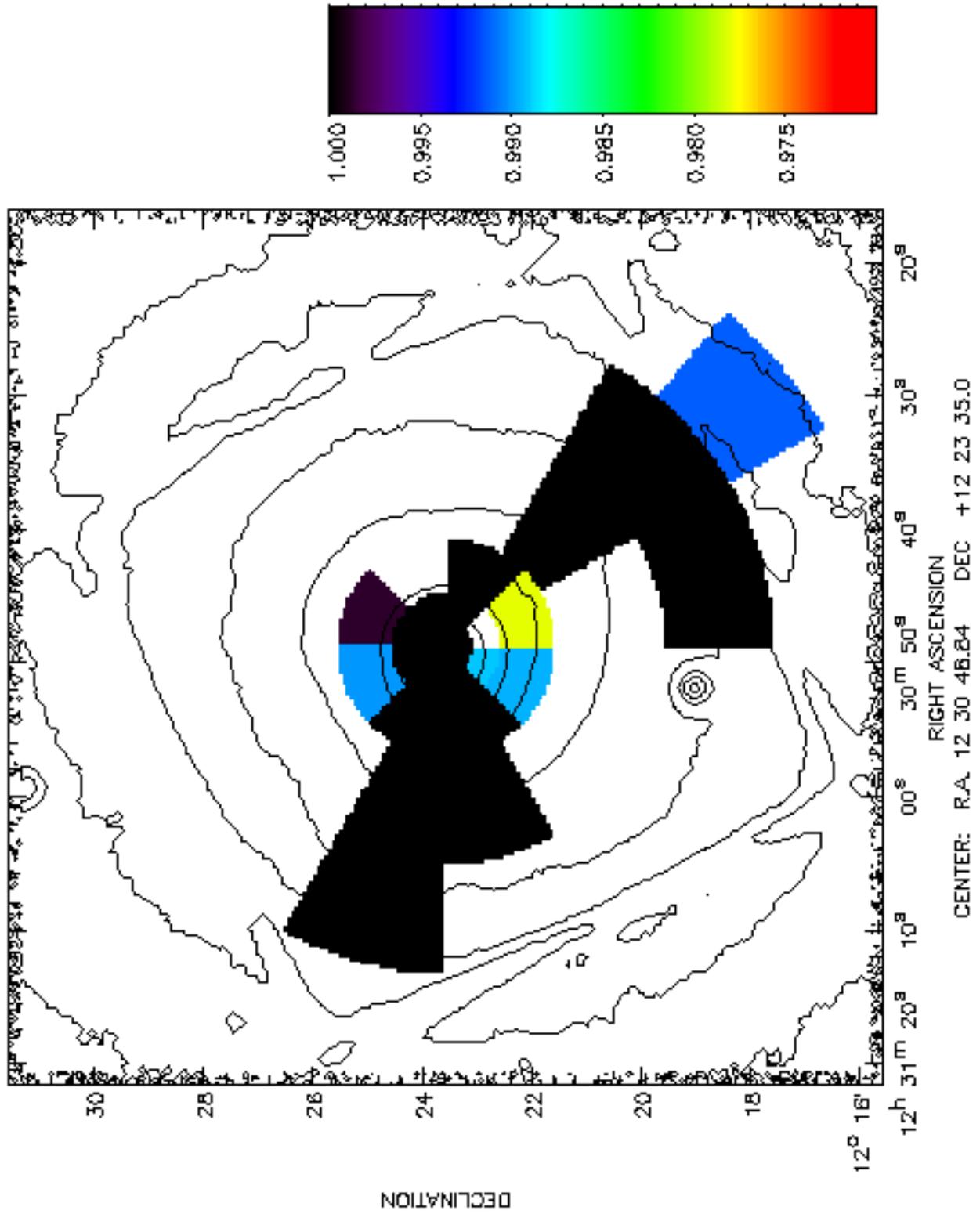}
\caption
{F-Probability map obtained by comparing the best fits of the 1 and 2 
temperature runs, overlaid on intensity contours. The color palette on the
 right shows the correspondence between colors and probability values.}
\end{figure}
\clearpage

\begin{figure}
\epsscale{1.0}
\plotone{f3.eps}
\caption
{Ratio of the 90\% confidence upper limit for the normalization of the third 
temperature component over the normalization of the cool component, as 
determined from the 2 temperature fit, versus the temperature of the third component.
The spectrum used for this exercise is from  the sector with bounding radii 
2$^\prime$ and 3$^\prime$ and bounding angles 150$^o$ and 180$^o$ 
(angles are counted starting from West and going counter clockwise).}
\end{figure}
\clearpage

\begin{figure}
\epsscale{0.80}
\plotone{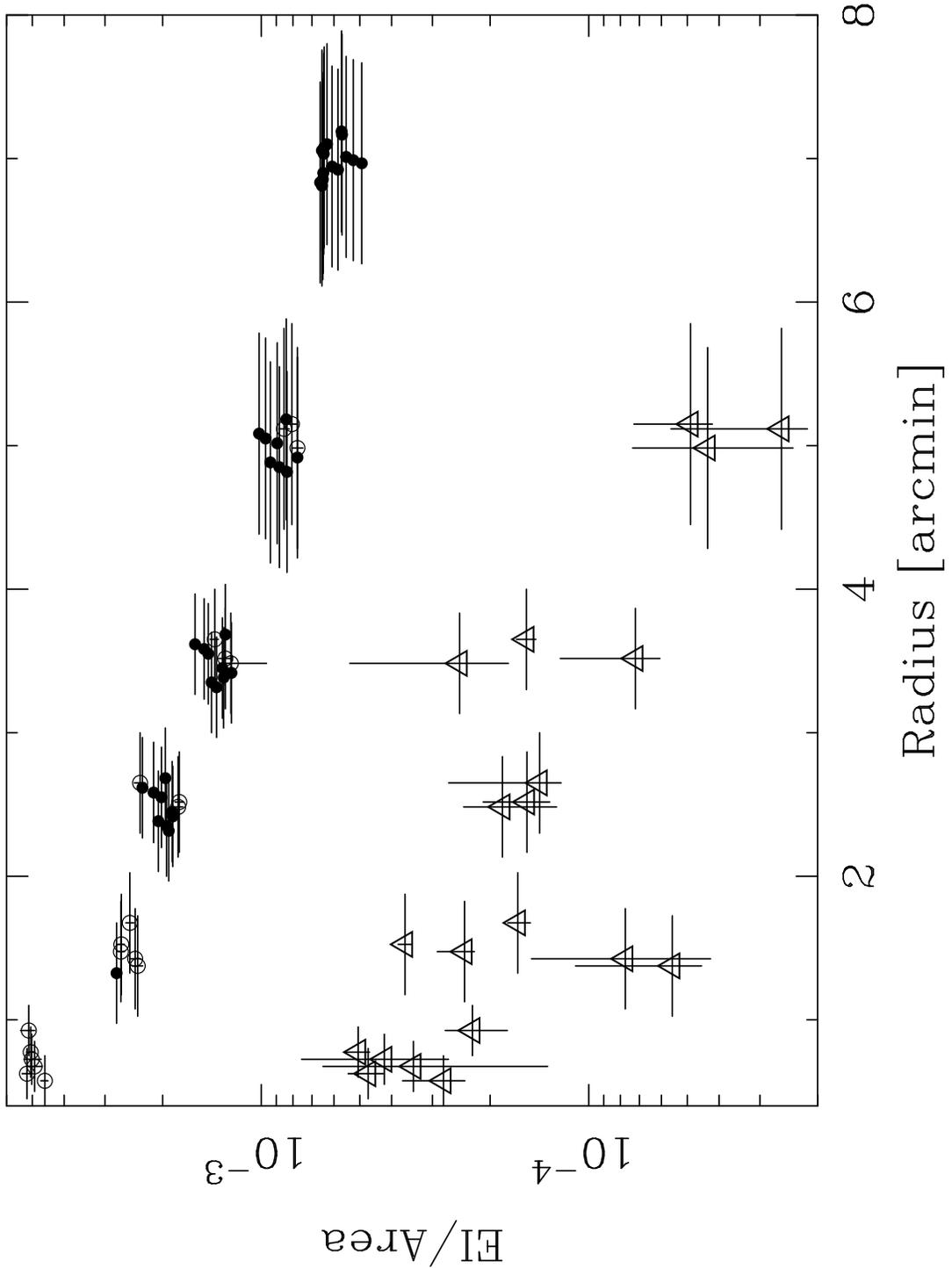}
\caption
{Normalized Emission Integral, i.e. EI/Area, versus radius. EI $\equiv 
{10^{-14} \over 4 \pi d_A^2(1+z)^2} \int n_e n_H dV$, where $d_A$ is the angular
distance to the source in cm, $z$ is the redshift, $n_e$ is the electron density 
in cm$^{-3}$, and $n_H$ is the hydrogen density in cm$^{-3}$. Area is in units of 
arcmin$^2$. Full circles are measurements from 1 temperature regions, 
empty circles and empty triangles indicate respectively measurements for the 
hot and cool components in 2 temperature regions.}
\end{figure}
\clearpage

\begin{figure}
\epsscale{0.85}
\plotone{f5.eps}
\caption
{Temperature versus radius. Symbols as in Fig. 4.}
\end{figure}
\clearpage

\begin{figure}
\epsscale{1.0}
\plotone{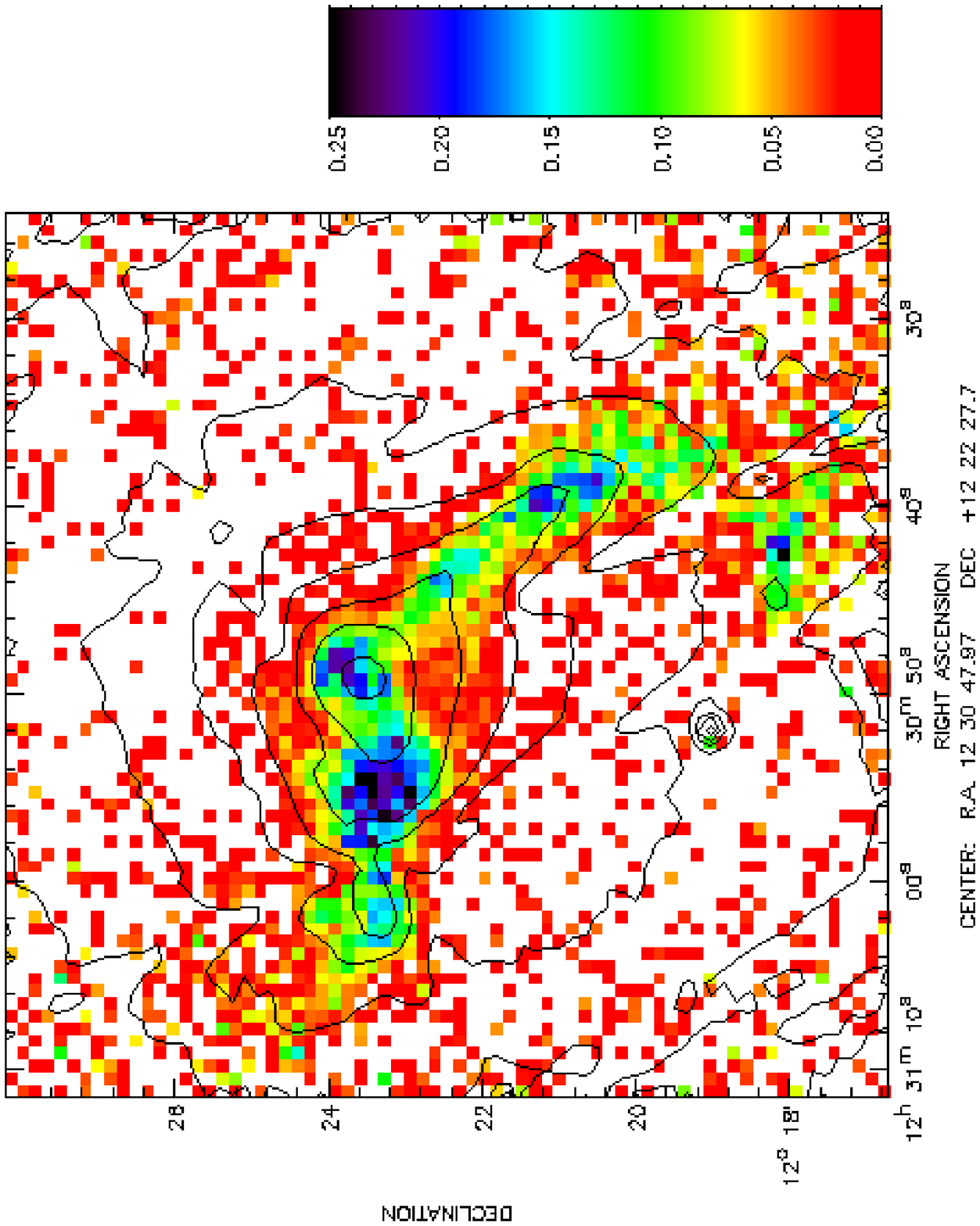}
\caption
{Emission integral ratio map, derived from EPIC data, overlaid on intensity 
contours, pixels are $10^{\prime\prime}\times10^{\prime\prime}$ or
750pc$\times$750pc.
The color palette on the right shows the correspondence between colors and 
Emission Integral ratios.}
\end{figure}
\clearpage

\begin{figure}
\epsscale{1.0}
\plotone{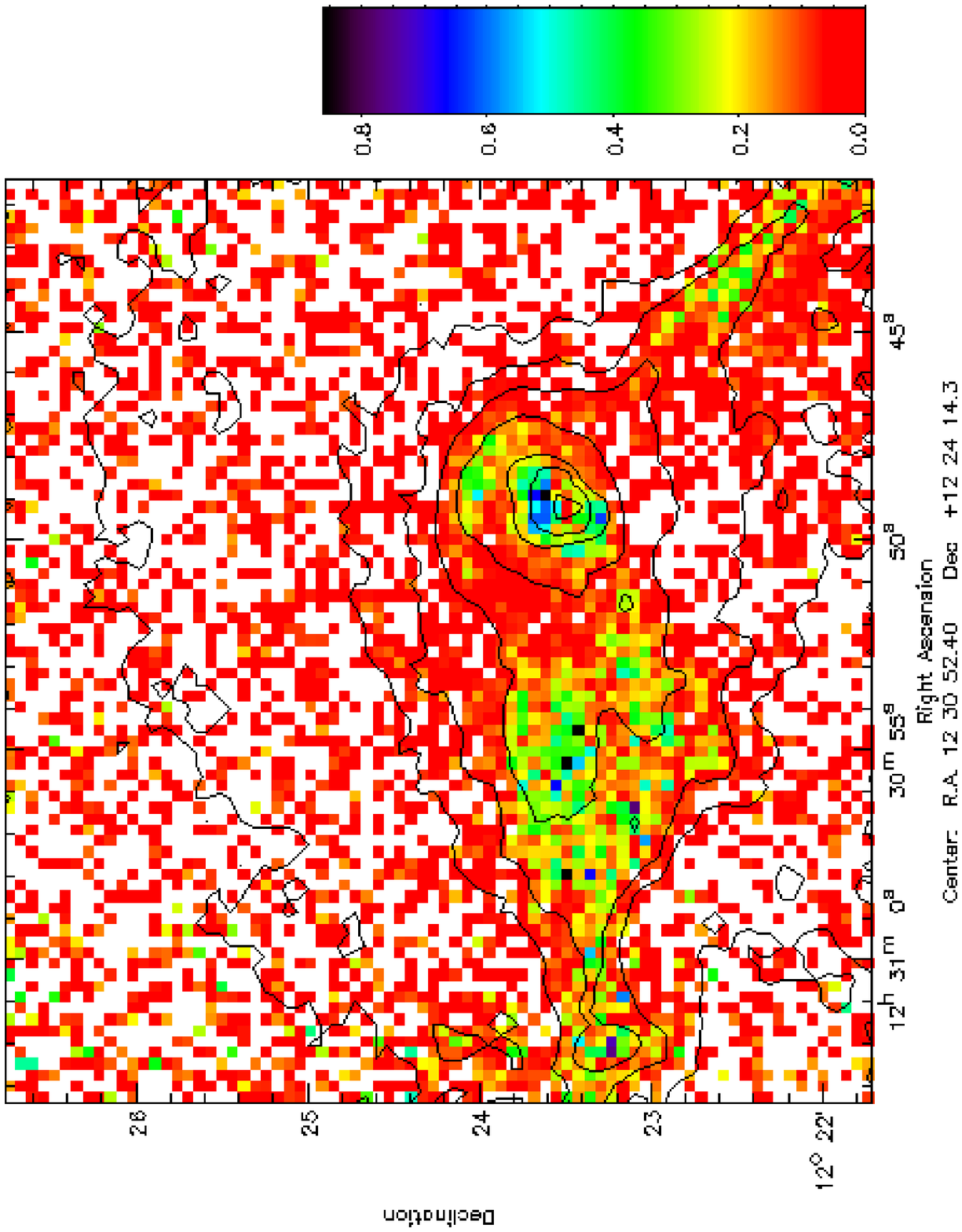}
\caption
{Emission integral ratio map, derived from Chandra data, overlaid on intensity 
contour, pixels are $4^{\prime\prime}\times4^{\prime\prime}$ or 
300pc$\times$300pc.
The color palette on the right shows the correspondence between colors and 
Emission Integral ratios.}
\end{figure}
\clearpage

\begin{figure}
\epsscale{0.85}
\plotone{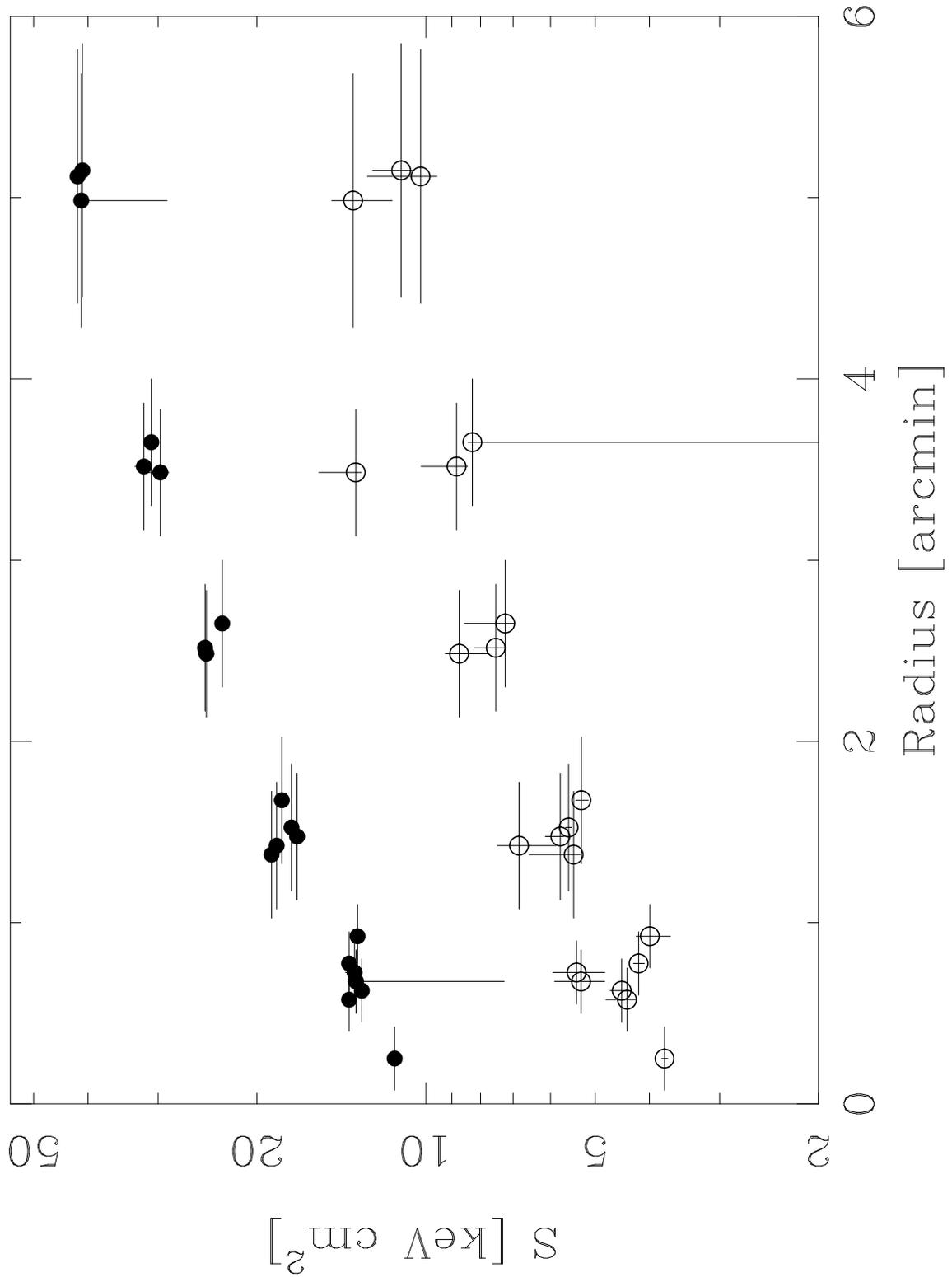}
\caption
{Entropy, defined as $S \equiv kT/n^{2/3}$,  versus radius for 2 temperature 
regions. Full and empty circles indicate respectively measurements for the 
hot and cool component.}
\end{figure}
\clearpage
 
\end{document}